\documentclass[11pt,twocolumn]{article}
\title{\textbf{\fontsize{16}{16}\selectfont Observational constraints on the fractal cosmology}}

\usepackage{geometry}
\usepackage{amsmath,amssymb,amsbsy}
\usepackage{graphicx}
\usepackage{epstopdf}
\usepackage{setspace}
\usepackage{flushend}
\usepackage[T1]{fontenc}
\usepackage{mathptmx}
\usepackage[numbers,sort&compress]{natbib}
\usepackage[colorlinks,citecolor=blue,urlcolor=blue,linkcolor=blue]{hyperref}

\usepackage{sectsty}
\sectionfont{\fontsize{11}{10}\selectfont}
\usepackage[font=footnotesize,labelfont=bf]{caption}
\usepackage{authblk}
\author[1,2]{Mahnaz Asghari \thanks{mahnaz.asghari@shirazu.ac.ir}}
\author[1,2]{Ahmad Sheykhi \thanks{asheykhi@shirazu.ac.ir}}
\affil[1]{\textit{\small Department of Physics, College of Sciences, Shiraz University, Shiraz 71454, Iran}}
\affil[2]{\textit{\small Biruni Observatory, College of Sciences, Shiraz University, Shiraz 71454, Iran}}
\date{}

\begin{document}
\maketitle
\begin{abstract}
In this paper, we explore a fractal model of the universe proposed
by Calcagni [JHEP{\bf03}(2010)120] for a power-counting
renormalizable field theory living in a fractal spacetime.
Considering a timelike fractal profile, we derived field 
equations in fractal cosmology, in order to explore the structure 
formation and the expansion history in fractal universe. 
Numerical investigations based on matter power spectra diagrams 
report higher structure growth in fractal cosmology, being in 
contrast to local galaxy surveys. Additionally, according to 
the evolution of Hubble parameter diagrams, it can be understood 
that Hubble constant would decrease in fractal cosmology, 
which is also incompatible with low redshift estimations of $H_0$.
So, concerning primary numerical studies, it seems that fractal 
cosmology is not capable to alleviate the tensions between 
local and global observational probes. 
Then, in pursuance of more accurate results, we constrain 
the fractal cosmology by observational data, including 
Planck cosmic microwave background (CMB), weak lensing, 
supernovae, baryon acoustic oscillations (BAO), 
and redshift-space distortions (RSD) data. 
The derived constraints on fractal dimension $\beta$ 
indicate that there is no considerable deviation from 
standard model of cosmology.
\end{abstract}

\section{Introduction}
Recent observational data from type Ia supernovae (SNeIa)
\cite{sn1,sn2}, the cosmic microwave background (CMB) anisotropies
\cite{cmb1,cmb2,cmb3}, large scale structures
\cite{lss1,lss2,lss3}, and baryon acoustic oscillations (BAO)
\cite{bao1,bao2,bao3}, confirm $\mathrm{\Lambda}$CDM as the concordance
model of cosmology. On the other hand, $\mathrm{\Lambda}$CDM model which is
based on the theory of general relativity (GR), encounters some
observational discrepancies, principally $\sigma_8$ and $H_0$
tensions. To be more specific, there is a disagreement between
low-redshift determinations and Planck CMB measurements of matter
perturbation amplitude, $\sigma_8$ \cite{s8,cmb3}. Moreover,
direct measurements of Hubble constant are in significant tension
with Planck observations \cite{H01,H02,H03,H04,cmb3}. So,
inconsistencies between local and global data, motivate
cosmologists to prob beyond $\mathrm{\Lambda}$CDM model. Accordingly, it is
suggested to consider some corrections on GR, describing as
modified theory of gravity \cite{mg1,mg2,mg3,mg4}.

On the other hand, Calcagni proposed an effective quantum field
theory which is power counting renormalizable and Lorentz
invariant, living in a fractal universe \cite{cal1,cal2}. The
fractal nature firstly introduced by Mandelbrot in 1983
\cite{man1} suggests conditional cosmological principle in fractal
universe, where the universe appears the same from every galaxy.
Thereafter, in 1986, Linde \cite{linde} propounded a model of an
eternally existing chaotic inflationary universe, explaining a
fractal cosmology. Accordingly, there are several investigations
on the theory of fractal cosmology in literatures. Rassem and
Ahmed \cite{thf1} in 1996 considered a nonhomogeneous cosmological
model with a fractal distribution of matter which evolves to a
homogeneous universe as time passes. The conditional cosmological
principle in fractal cosmology is discussed in \cite{thf2,thf3}.
\cite{thf4} studies Multi-fractal geometry. Thermodynamics of the
apparent horizon in a fractal universe is explored in \cite{thf5}.
In order to find more theoretical studies on fractal cosmology
refer to e.g. \cite{thf6,thf7,thf8,thf9,thf10}. Furthermore, it is
interesting to explore fractal models with cosmological data as
discussed in \cite{obf1,obf2,obf3,obf4,obf5,obf6} (also see
\cite{obf7} for a review). Correspondingly, in the present work we
are going to investigate the fractal universe in background and
perturbation levels, as well as studying observational constraints
on parameters of fractal cosmology.

The paper is organized as follows. Section \ref{sec2} is dedicated
to field equations in a fractal universe. In section \ref{sec3},
we study the fractal model numerically, and further we constrain
the model with current observational data in section \ref{sec4}.
We summarize our results in section \ref{sec5}.
\section{Field equations in a fractal universe} \label{sec2}
The total action in a fractal spacetime is given by \cite{cal1}
\begin{align} \label{eq1}
S=\frac{1}{16\pi G}\int{\mathrm{d}\mathcal{\varrho}(x)\,
\sqrt{-g}\big(R-2\mathrm{\Lambda}-\omega\partial_{\mu}v\partial^{\mu}v\big)}+S_m
\,,
\end{align}
where $\omega$ is the fractal parameter, $v$ is the fractional
function, and $\mathrm{d}{\varrho}(x)$ is a Lebesgue-Stieltjes
measure. It is possible to derive field equations from action
(\ref{eq1}) similar to scalar-tensor theories. Thus, in a fractal
universe we obtain \cite{cal1}
\begin{align} \label{eq2}
& R_{\mu \nu}-\frac{1}{2}g_{\mu \nu}(R-2\mathrm{\Lambda}) +g_{\mu
\nu}\frac{\Box{v}}{v}-\frac{\nabla_{\mu} \nabla_{\nu}
v}{v} \nonumber \\
&+\omega\Big(\frac{1}{2}g_{\mu
\nu}\partial_{\sigma}v\partial^{\sigma}v-\partial_{\mu}v\partial_{\nu}v\Big)=8\pi
G T_{\mu \nu} \,.
\end{align}
Furthermore, continuity equation in a fractal spacetime takes the form \cite{cal1}
\begin{align} \label{eq3}
\nabla_{\mu}\big(vT^{\mu}_{\nu}\big)-\partial_{\nu}v
\mathcal{L}_m=0 \,.
\end{align}
It should be noted that for $v=1$, standard equations in GR
will be recovered. 
Here, we focus on a timelike fractal, then $v$ is only time dependent given by
\begin{equation} \label{eq5}
	v=H_0^{\beta}\Big(\frac{a}{a_0}\Big)^{\beta} \,,
\end{equation}
in which $\beta=4(1-\alpha)$ is the fractal dimension, and the
parameter $\alpha$ ranges as $0<\alpha\leq1$.

We consider a fractal universe with
the following flat Friedmann-Lema\^itre-Robertson-Walker (FLRW)
metric in the synchronous gauge
\begin{equation} \label{eq4}
\mathrm{d}s^2=a^2(\tau)\Big(-\mathrm{d}\tau^2+\big(\delta_{ij}+h_{ij}\big)\mathrm{d}x^i\mathrm{d}x^j\Big) \;,
\end{equation}
in which 
\begin{equation*}
h_{ij}(\vec{x},\tau)=\int \mathrm{d}^3k\,e^{i\vec{k}.\vec{x}}
\bigg(\hat{k}_i\hat{k}_jh(\vec{k},\tau)+\Big(\hat{k}_i\hat{k}_j-\frac{1}{3}\delta_{ij}\Big)6\eta(\vec{k},\tau)\bigg) ,
\end{equation*} 
with scalar perturbations $h$ and $\eta$, and $\vec{k}=k\hat{k}$ \cite{pt}.
Then, considering the energy content of the universe as a perfect fluid 
with $T_{\mu\nu}=\big(\rho+p\big)u_{\mu}u_{\nu}+g_{\mu \nu}p$, 
field equations in background level take the form
\begin{align}
& H^2\bigg(1+\beta-\frac{1}{6}\omega H_0^{2\beta}\beta^2 \Big(\frac{a}{a_0}\Big)^{2\beta}\bigg)=\frac{8\pi G}{3}\sum_{i}\bar{\rho}_i \,, \label{eq6} \\
& (\beta+2)\frac{H'}{a}+H^2\bigg(3+2\beta+\beta^2+\frac{1}{2}\omega H_0^{2\beta}\beta^2 \Big(\frac{a}{a_0}\Big)^{2\beta}\bigg) \nonumber \\
&=-8\pi G \sum_{i}\bar{p}_i \,, \label{eq7}
\end{align}
where a prime indicates a deviation with respect to the conformal
time. It can be easily seen that, $\beta=0$ restores field
equations in standard cosmology. According to equation
(\ref{eq6}), total density parameter can be find as
\begin{equation} \label{eq8}
\mathrm{\Omega}_\mathrm{tot}=1+\beta-\frac{1}{6}\omega H_0^{2\beta}\beta^2 \Big(\frac{a}{a_0}\Big)^{2\beta} \,,
\end{equation}
where we have considered a universe filled with radiation (R),
baryons (B), dark matter (DM) and cosmological constant
($\mathrm{\Lambda}$). Also field equations to linear order of perturbations
can be written as
\begin{align}
& \frac{a'}{a}\Big(1+\frac{1}{2}\beta\Big)h'-2k^2\eta=8\pi G a^2 \sum_{i}\delta \rho_{i} \,, \label{eq20} \\
& k^2\eta'=4\pi G a^2 \sum_{i}\big(\bar{\rho}_i+\bar{p}_i\big)\theta_{i} \,, \label{eq21} \\
& \frac{1}{2}h''+3\eta''+\frac{a'}{a}\big(h'+6\eta'\big)\Big(1+\frac{1}{2}\beta\Big)-k^2\eta=0 \,, \label{eq22} \\
& \frac{a'}{a}\Big(2+\beta\Big)h'+h''-2k^2\eta=-24\pi G a^2 \sum_{i}\delta p_{i}  \,. \label{eq23}
\end{align}
In addition, regarding equation (\ref{eq3}), conservation
equations of fractal cosmology for $i$th component of the universe
in background and perturbation levels become
\begin{align}
\bar{\rho}'_i+\big(3+\beta\big)\frac{a'}{a}\big(\bar{\rho}_i+\bar{p}_i\big)=0 \,, \label{eq28} 
\end{align}
\begin{align}
\delta'_{i}=&-\frac{a'}{a}\big(3+\beta\big)\bigg[\delta_{i}\big(c^2_{si}-w_i\big) \nonumber \\
&+\big(c^2_{si}-c^2_{ai}\big)\frac{a'}{a}\big(3+\beta\big)\big(1+w_i\big)\frac{\theta_{i}}{k^2}\bigg] \nonumber \\
&-\big(1+w_i\big)\theta_{i}-\frac{1}{2}\big(1+w_i\big)h' \,, \label{eq29} 
\end{align}
\begin{align}
\theta'_{i}=&\theta_{i}\bigg[-\frac{a'}{a}\big(4+\beta\big)+\frac{a'}{a}\big(3+\beta\big)
\big(1+w_i+c^2_{si}-c^2_{ai}\big)\bigg] \nonumber \\
&+\frac{k^2c^2_{si}}{1+w_i}\delta_{i} \,. \label{eq30}
\end{align}
Then, choosing $\beta=0$ would recover equations in concordance
$\mathrm{\Lambda}$CDM model.

Now that we have described main equations in fractal cosmology, it
is possible to derive observational constraints on fractal model,
using a modified version of the CLASS\footnote{Cosmic Linear
Anisotropy Solving System} code \cite{cl}, and also applying an
MCMC\footnote{Markov Chain Monte Carlo} approach via the
M\textsc{onte} P\textsc{ython} code \cite{mp1,mp2}.
\section{Numerical results} \label{sec3}
In this section, we modify the CLASS cosmological Boltzmann code
according to the field equations of fractal model outlined in
section \ref{sec2}. Correspondingly, we take into account the
Planck 2018 data \cite{cmb3} (i.e.
$\mathrm{\Omega}_{\mathrm{B},0}h^2=0.02242$,
$\mathrm{\Omega}_{\mathrm{DM},0}h^2=0.11933$,
$H_0=67.66\,\mathrm{km\,s^{-1}\,Mpc^{-1}}$, $A_s=2.105\times
10^{-9}$, and $\tau_\mathrm{reio}=0.0561$) in our numerical
investigation. 

\begin{figure*}
    \includegraphics[width=9cm]{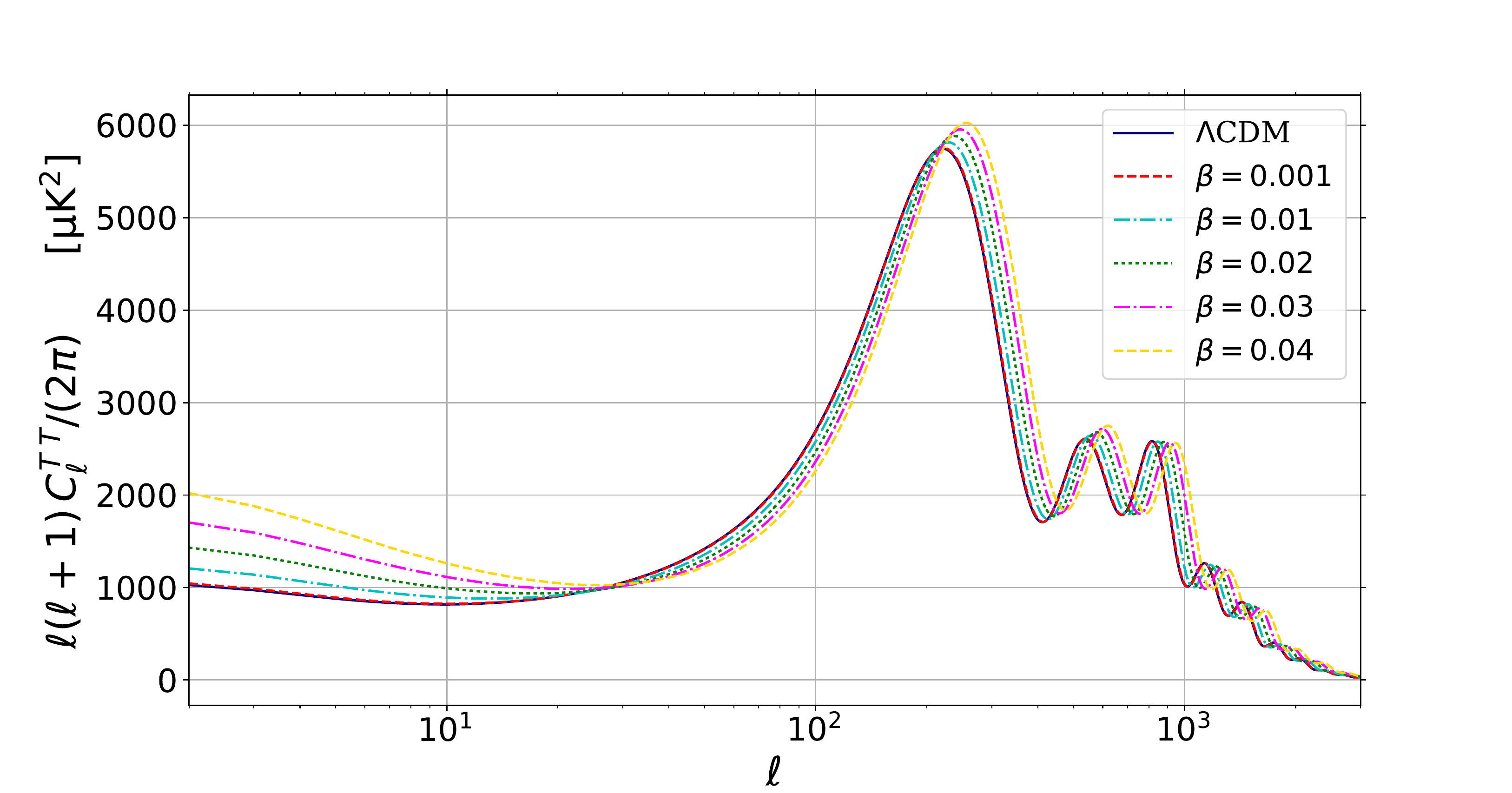}
    \includegraphics[width=9cm]{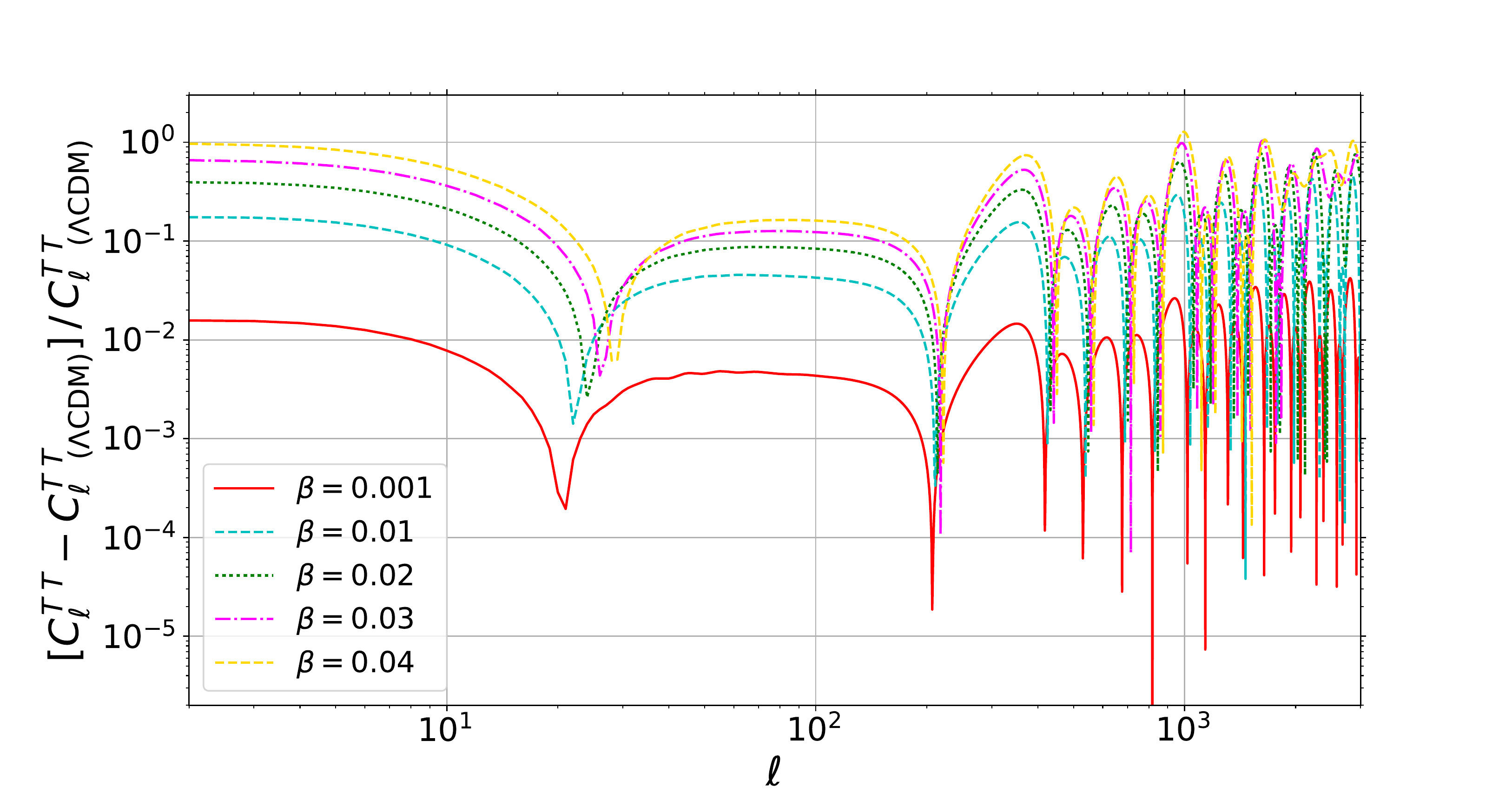}
    \includegraphics[width=9cm]{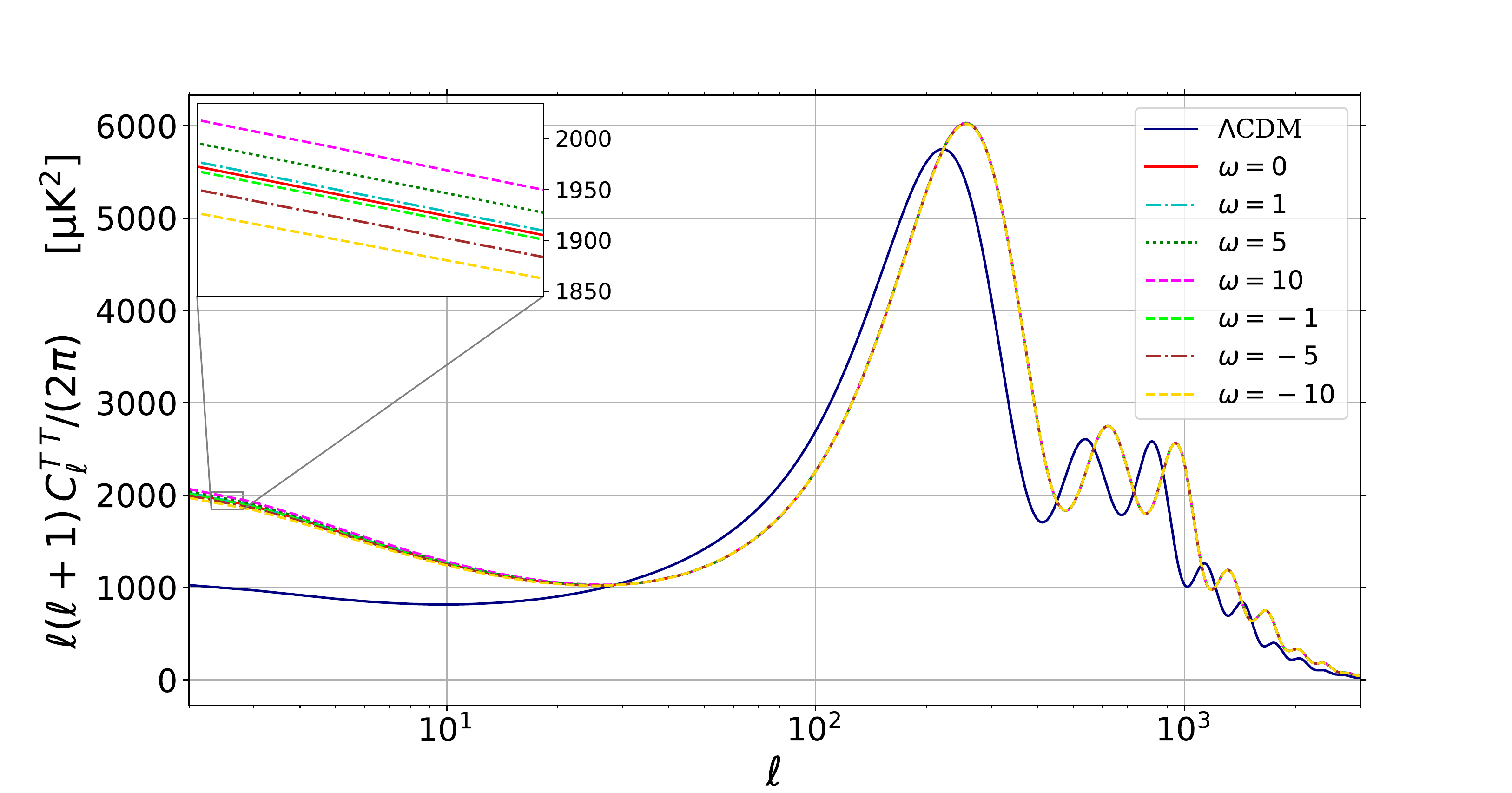}
    \includegraphics[width=9cm]{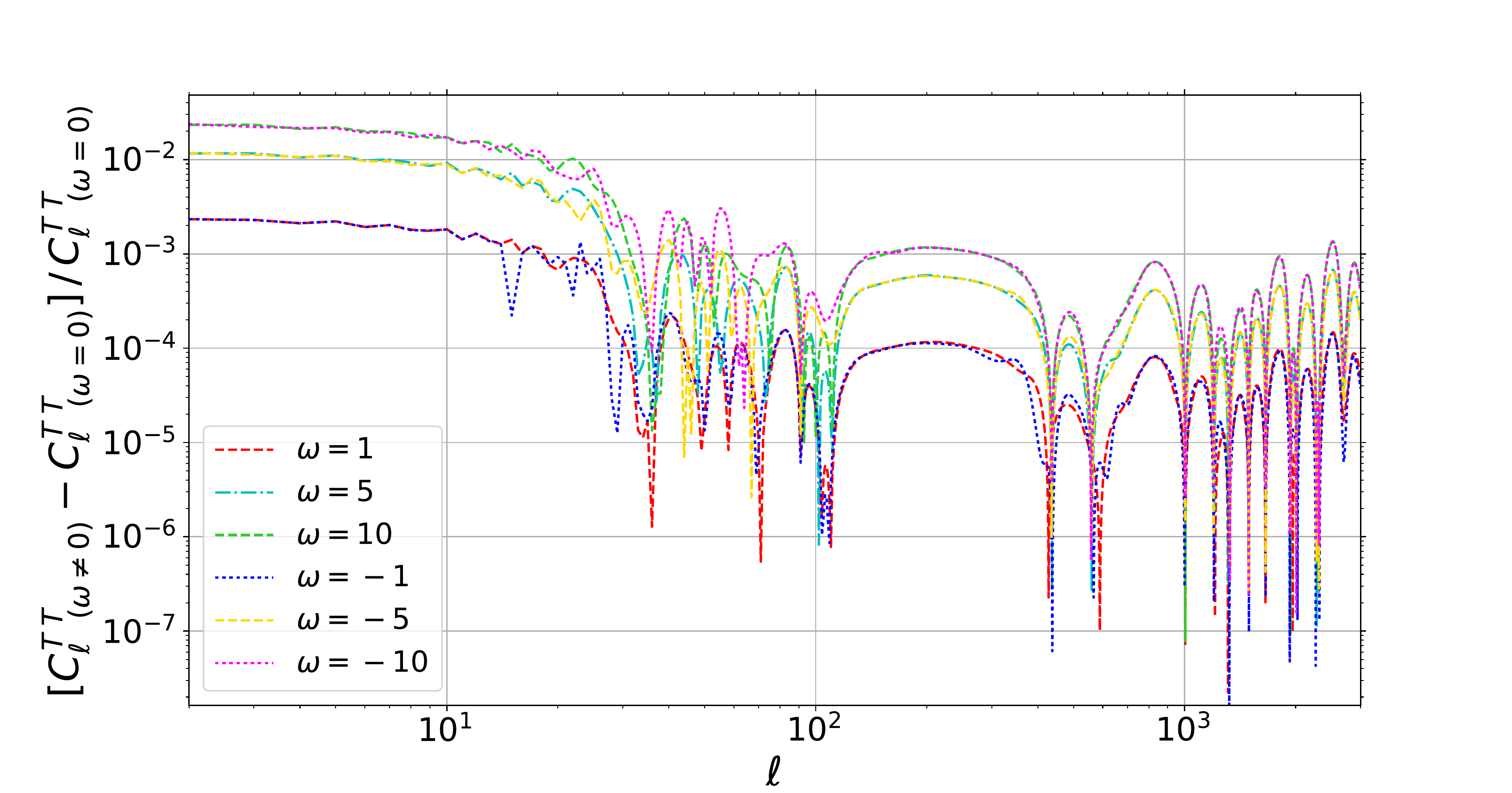}
    \caption{ Upper panels show the CMB power spectra diagrams (left) and their relative ratio with respect to standard cosmological model (right) for different values of $\beta$, considering $\omega=0$. Lower panels show the the CMB power spectra diagrams for different values of $\omega$, considering $\beta=0.04$, compared to $\mathrm{\Lambda}$CDM model (left) and their relative ratio with respect to the case $\omega=0$ (right).}
    \label{f1}
\end{figure*}

In Fig. \ref{f1} we depict the CMB temperature
power spectra in fractal cosmology compared to $\mathrm{\Lambda}$CDM model.
Upper panels display the CMB power spectra for different values of
$\beta$, where we have considered $\omega=0$. In lower panels the
CMB temperature anisotropy diagrams are illustrated for different
values of $\omega$, while regarding $\beta=0.04$. Furthermore, for
a better comparison we have shown the relative ratio with respect
to $\omega=0$ in the bottom-right panel, which indicates that the
effect of $\omega$ is not significant in CMB power spectra.

\begin{figure*}
    \includegraphics[width=9cm]{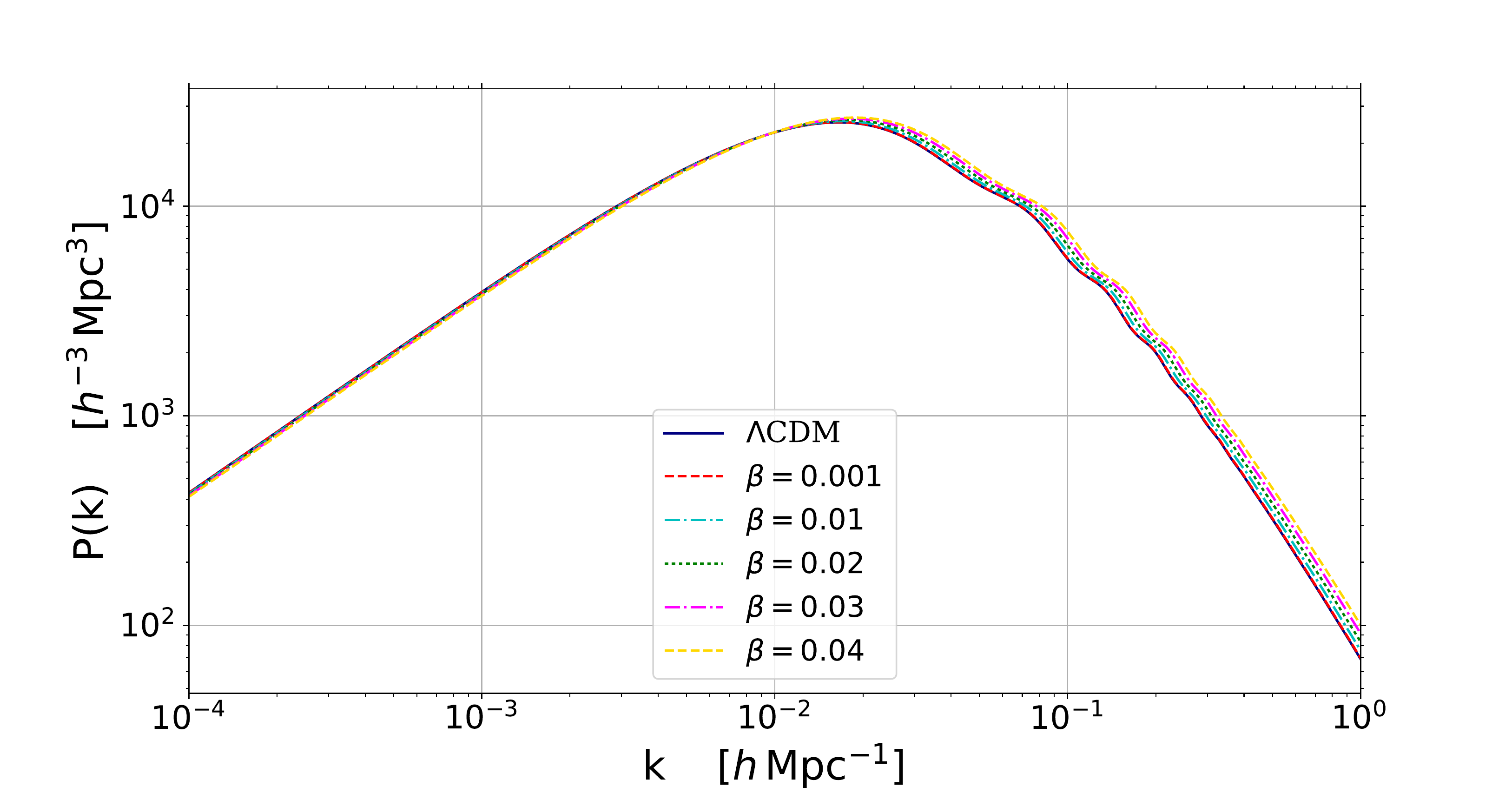}
    \includegraphics[width=9cm]{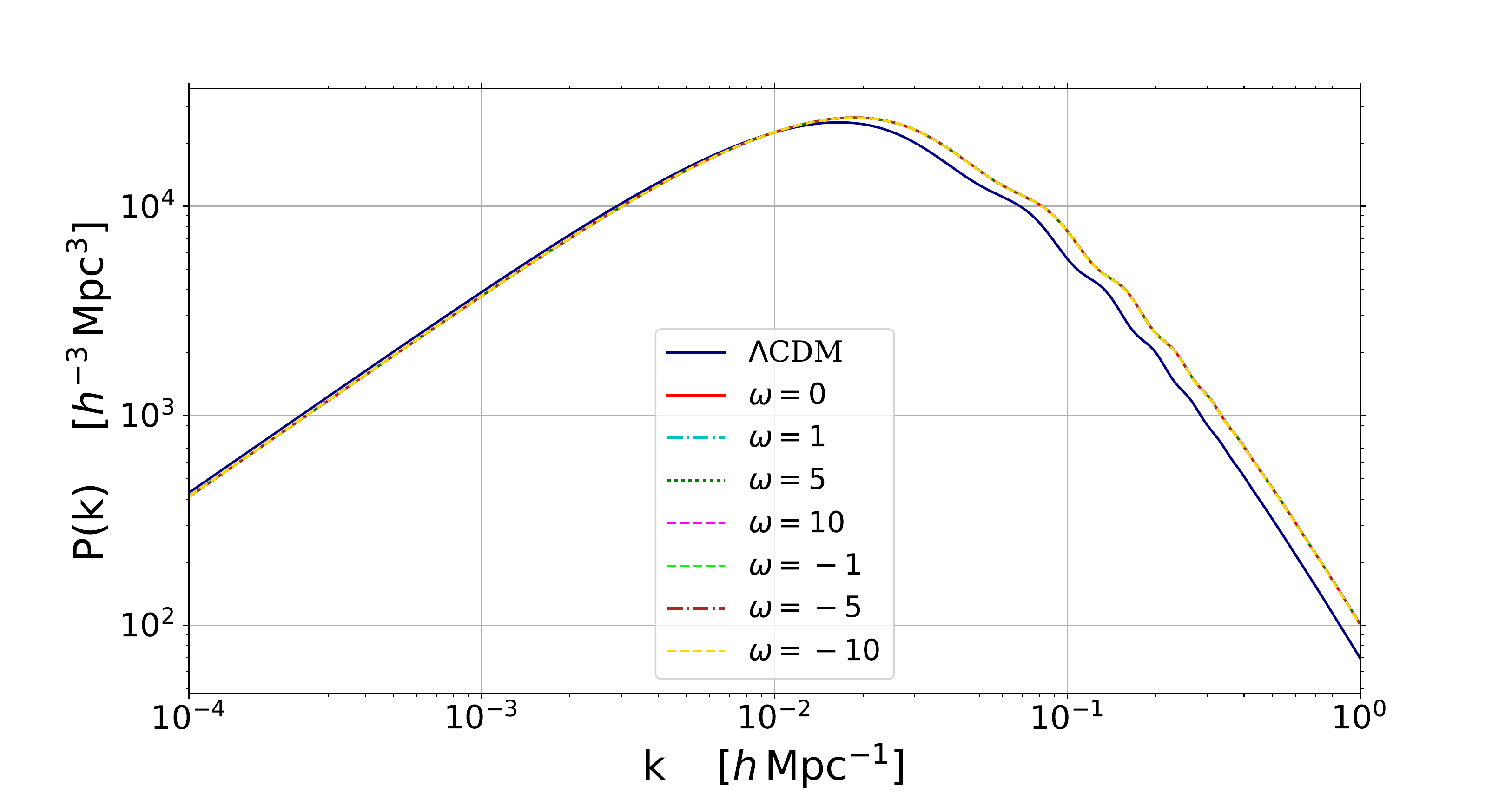}
    \caption{ Matter power spectra diagrams for different values of $\beta$ compared to $\mathrm{\Lambda}$CDM model, considering $\omega=0$ (left), and analogous diagrams for different values of $\omega$, where $\beta=0.04$ (right).}
    \label{f2}
\end{figure*}

Figure \ref{f2} demonstrates the matter power spectra of fractal
model in comparison with standard cosmological model. In left
panel we see matter power spectra for different values of $\beta$,
where $\omega$ is considered to be zero. Right panel shows the
power spectra for different values of $\omega$, in which we have
fixed $\beta$ to be $0.04$. Considering this figure, fractal
cosmology predicts a higher growth of structure which is in
conflict with local probes of large scale structures. In addition,
this plot explains that the fractal parameter $\omega$ has no
remarkable influence on structure formation.

\begin{figure*}
    \includegraphics[width=9cm]{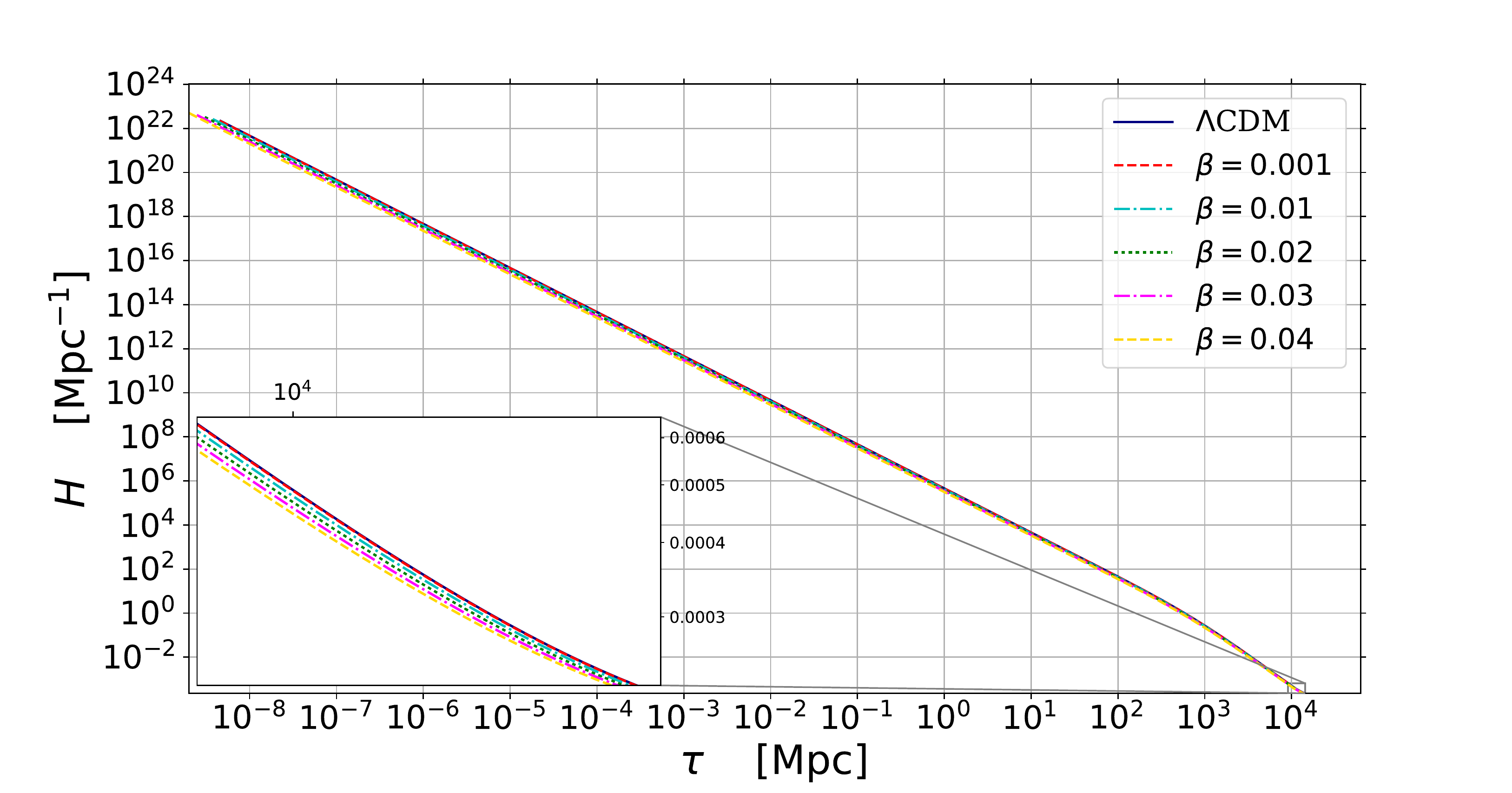}
    \includegraphics[width=9cm]{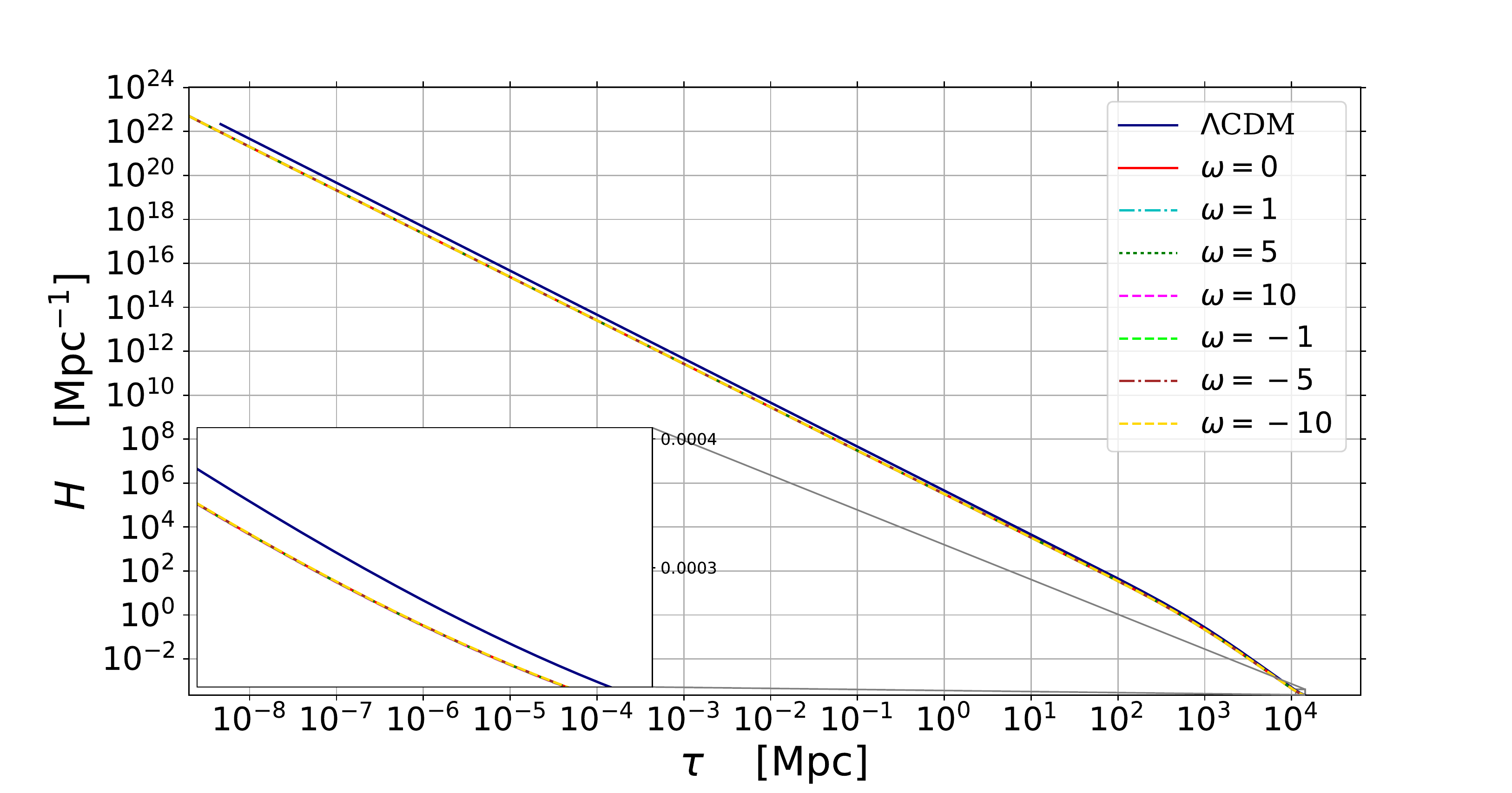}
    \caption{ Hubble parameter in term of conformal time for different values of $\beta$ compared to $\mathrm{\Lambda}$CDM model, where $\omega=0$ (left), and analogous diagrams for different values of $\omega$, regarding $\beta=0.04$ (right).}
    \label{f3}
\end{figure*}

Moreover, the evolution of Hubble parameter in fractal cosmology
is depicted in Fig. \ref{f3}. Accordingly, we can conclude
that Hubble tension is aggravated in fractal model, and also the
expansion history would not be affected by the fractal parameter
$\omega$.
\section{Constraints from observational data} \label{sec4}
In this part, we employ the MCMC sampling package M\textsc{onte}
P\textsc{ython} to investigate observational constraints on
fractal model. In this respect, we use the following set of
cosmological parameters: \{ $100\,\mathrm{\Omega}_{\mathrm{B},0} h^2$,
$\mathrm{\Omega}_{\mathrm{DM},0} h^2$, $100\,\theta_s$, $\ln (10^{10}
A_s)$, $n_s$, $\tau_{\mathrm{reio}}$, $\beta$ \}, \\
where $\mathrm{\Omega}_{\mathrm{B},0} h^2$ and $\mathrm{\Omega}_{\mathrm{DM},0} h^2$
stand for the baryon and cold dark matter densities respectively,
$\theta_s$ represents the ratio of the sound horizon to the
angular diameter distance at decoupling, $A_s$ indicates the
amplitude of the primordial scalar perturbation spectrum, $n_s$
stands for the scalar spectral index, $\tau_{\mathrm{reio}}$ is
the optical depth to reionization, and $\beta$ is the fractal
dimension. We have also put constraints on four derived parameters
which are reionization redshift ($z_\mathrm{reio}$), the matter
density parameter ($\mathrm{\Omega}_{\mathrm{M},0}$), the Hubble constant
($H_0$), and the root-mean-square mass fluctuations on scales of 8
$h^{-1}$ Mpc ($\sigma_8$). Additionally, preliminary numerical
explorations consider the prior range [$0$, $0.04$] for the
fractal dimension. It is worth mentioning that, since the fractal
parameter $\omega$ is irrelevant to cosmological observable (based
on numerical results in section \ref{sec3}), we fix $\omega$ to be
zero in MCMC analysis.

The dataset we use to constrain fractal model includes the Planck
likelihood with Planck 2018 data (containing high-$l$ TT,TE,EE,
low-$l$ EE, low-$l$ TT, and lensing) \cite{cmb3}, the Planck-SZ
likelihood for the Sunyaev-Zeldovich effect measured by Planck
\cite{sz1,sz2}, the CFHTLenS likelihood with the weak lensing data
\cite{lens1,lens2}, the Pantheon likelihood with the supernovae
data \cite{pan},  the BAO likelihood with the baryon acoustic
oscillations data \cite{bao4,bao5}, and the BAORSD likelihood for
BAO and redshift-space distortions (RSD) measurements
\cite{rsd1,rsd2}.

\begin{figure}
    \centering
    \includegraphics[width=8.5cm]{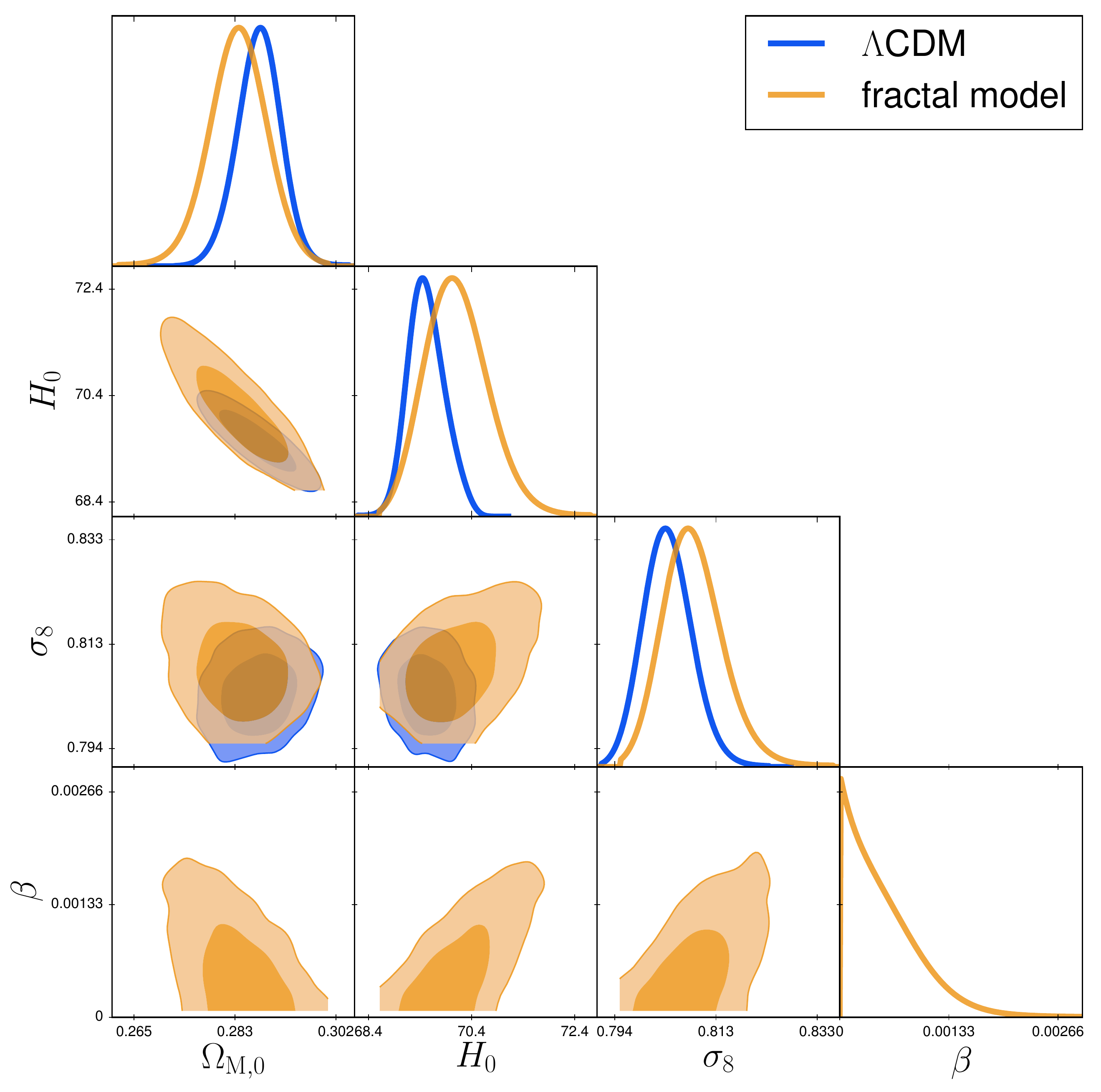}
    \caption{ The one-dimensional posterior distribution and two-dimensional posterior contours with 68\% and 95\% confidence limits for the selected cosmological parameters of fractal model (orange) compared to $\mathrm{\Lambda}$CDM (blue).}
    \label{f4}
\end{figure}

The best fit and $1\sigma$ and $2\sigma$ confidence intervals for
cosmological parameters of fractal model and also $\mathrm{\Lambda}$CDM as
the reference model, from "Planck + Planck-SZ + CFHTLenS +
Pantheon + BAO + BAORSD" dataset are summarized in Table
\ref{t1}.
\begin{table}
    \centering
    \caption{ Best fit values and 68\% and 95\% confidence limits for cosmological parameters from "Planck + Planck-SZ +
        CFHTLenS + Pantheon + BAO + BAORSD" data set for $\mathrm{\Lambda}$CDM and fractal model.}
    \label{t1}
    \scalebox{.6}{
        \begin{tabular}{lllll}
            \cline{1-5}
            & \multicolumn{2}{c}{} & \multicolumn{2}{c}{} \\
            & \multicolumn{2}{c}{$\mathrm{\Lambda}$CDM} & \multicolumn{2}{c}{fractal model} \\
            \cline{2-5}
            & & & & \\
            {parameter} & best fit & 68\% \& 95\% limits & best fit & 68\% \& 95\% limits \\ \hline
            & & & & \\
            $100\,\mathrm{\Omega}_{\mathrm{B},0} h^2$ & $2.261$ & $2.263^{+0.012+0.026}_{-0.013-0.025}$ & $2.274$ & $2.265^{+0.014+0.028}_{-0.015-0.027}$ \\
            & & & & \\
            $\mathrm{\Omega}_{\mathrm{DM},0} h^2$ & $0.1163$ & $0.1164^{+0.00078+0.0015}_{-0.00079-0.0015}$ & $0.1164$ & $0.1168^{+0.00086+0.0017}_{-0.00093-0.0018}$ \\
            & & & & \\
            $100\,\theta_s$ & $1.042$ & $1.042^{+0.00029+0.00055}_{-0.00026-0.00053}$ & $1.042$ & $1.042^{+0.00030+0.00055}_{-0.00028-0.00054}$ \\
            & & & & \\
            $\ln (10^{10} A_s)$ & $3.034$ & $3.024^{+0.010+0.023}_{-0.014-0.021}$ & $3.023$ & $3.025^{+0.0091+0.022}_{-0.014-0.021}$ \\
            & & & & \\
            $n_s$ & $0.9712$ & $0.9719^{+0.0036+0.0072}_{-0.0039-0.0074}$ & $0.9724$ & $0.9720^{+0.0036+0.0074}_{-0.0038-0.0072}$ \\
            & & & & \\
            $\tau_\mathrm{reio}$ & $0.05358$ & $0.04963^{+0.0041+0.010}_{-0.0074-0.0096}$ & $0.04706$ & $0.04919^{+0.0039+0.010}_{-0.0077-0.0092}$ \\
            & & & & \\
            $\beta$ & --- & --- & $0.0003854$ & $0.0005421^{+0.00012+0.00092}_{-0.00054-0.00054}$ \\
            & & & & \\
            $z_\mathrm{reio}$ & $7.502$ & $7.084^{+0.50+1.0}_{-0.69-1.0}$ & $6.811$ & $7.046^{+0.49+1.0}_{-0.70-1.0}$ \\
            & & & & \\
            $\mathrm{\Omega}_{\mathrm{M},0}$ & $0.2871$ & $0.2876^{+0.0043+0.0086}_{-0.0044-0.0086}$ & $0.2828$ & $0.2841^{+0.0056+0.010}_{-0.0050-0.012}$ \\
            & & & & \\
            $H_0\;[\mathrm{km\,s^{-1}\,Mpc^{-1}}]$ & $69.56$ & $69.54^{+0.37+0.73}_{-0.36-0.71}$ & $70.16$ & $70.08^{+0.46+1.3}_{-0.68-1.1}$ \\
            & & & & \\
            $\sigma_8$ & $0.8079$ & $0.8044^{+0.0045+0.0096}_{-0.0051-0.0091}$ & $0.8063$ & $0.8092^{+0.0049+0.012}_{-0.0067-0.011}$ \\
            & & & & \\
            \cline{1-5}
        \end{tabular}
    }
\end{table}
Correspondingly, contour plots for some selected parameters of
fractal model compared to $\mathrm{\Lambda}$CDM are displayed in Fig.
\ref{f4}.
Considering the obtained constraints on fractal dimension,
observational data detect no significant deviation from
$\mathrm{\Lambda}$CDM model.

Moreover, the Akaike information criterion (AIC) defined as
$\mathrm{AIC}=-2\ln{\mathcal{L}_{\mathrm{max}}}+2K$ where
$\mathcal{L}_{\mathrm{max}}$ is the maximum likelihood function
and $K$ indicates the number of free parameters \cite{aic1,aic2},
results in $\mathrm{AIC_{(\mathrm{\Lambda} CDM)}}=3847.12$ and
$\mathrm{AIC_{(fractal)}}=3849.70$, then consequently
$\mathrm{\Delta AIC}=2.58$. Thus, $\mathrm{\Lambda}$CDM provides a better
fit to observations, while fractal model is still allowed.
\section{Conclusions} \label{sec5}
We have considered a fractal model of the universe introduced by
Calcagni \cite{cal1,cal2}, which is a power counting
renormalizable and also Lorentz invariant model. Taking into
account a timelike fractal, we have investigated the influence of
fractal model on observables, mainly power spectra and Hubble
constant, by using a modified version of the publicly available
CLASS code. Considering matter power spectra diagrams 
in Fig. \ref{f2}, there is an enhancement in structure growth 
for non-zero values of fractal dimension $\beta$,
being incompatible with low redshift structure formation.
Moreover, according to Fig. \ref{f3}, 
the current value of Hubble parameter decreases in
fractal cosmology, which is inconsistent with local determinations
of $H_0$. Consequently, primary numerical results indicate that fractal
cosmology has no positive impact on relieving cosmological
tensions.

Furthermore, we put constraints on the parameters of the fractal
universe by utilizing current observations, chiefly Planck CMB,
weak lensing, supernovae, BAO, and RSD data. 
Numerical results based on MCMC analysis, detect 
no significant departure from standard cosmological model. 
On the other hand, the model selection criterion AIC affirms that 
$\mathrm{\Lambda}$CDM model is more favored by observations, 
however, the fractal model can not be ruled out.
\section*{\fontsize{9}{9}\selectfont Acknowledgements
\, {\normalfont We thank Shiraz University Research Council. 
We are also grateful to the referee for valuable comments 
which helped us improve the paper significantly.}}
\section*{\fontsize{9}{9}\selectfont Data Availability Statement \, {\normalfont This manuscript has no associated data or the data
		will not be deposited.[Authors' comment: Observational data we have used in this paper are publicly available in Refs. \cite{cmb3,sz1,sz2,lens1,lens2,pan,bao4,bao5,rsd1,rsd2}. The modified version of the CLASS code is available under reasonable request.]}}
\bibliographystyle{unsrt}
\interlinepenalty=10000
\bibliography{1}

\end{document}